# Effect of Nd and Rh substitution on the spin dynamics of Kondo insulator $CeFe_2Al_{10}$

P. A. Alekseev[1, 2], J.-M. Mignot[3], D. T. Adroja[4,5], V. N. Lazukov[1], H. Tanida[6], Y. Muro[6], M. Sera[7], T. Takabatake[7], P. Steffens[8] and S. Rols[8]

[1]*National Research Centre "Kurchatov Institute" Moscow, 123182, Russian Federation*

[2]*National Research Nuclear University MEPhI, 115409, Moscow, Russian Federation*

[3]*Université Paris-Saclay, CNRS, CEA, Laboratoire Léon Brillouin, 91191 Gif sur Yvette cedex, France*

[4]*ISIS Facility, Rutherford Appleton Laboratory, Chilton, Didcot Oxon, OX11 0QX, United Kingdom*

[5]*Highly Correlated Matter Research Group, Physics Department, University of Johannesburg, PO Box 524, Auckland Park 2006, South Africa*

[6]*Center for Liberal Arts and Sciences, Faculty of Engineering, Toyama Prefectural University, Imizu 939-0398, Japan*

[7]*Graduate School of Advanced Science and Technology, Hiroshima University, Higashi-Hiroshima, Hiroshima 739-8530, Japan*

[8]*Institut Laue-Langevin, 71 avenue des Martyrs, CS 20156, 38042 Grenoble Cedex 9, France*

**Abstract**

The dynamic magnetic properties of the Kondo-insulator state in $CeFe_2Al_{10}$ (spin gap, resonance mode) have been investigated using polarized neutrons on a single crystal of pure $CeFe_2Al_{10}$. The results indicate that the magnetic excitations are polarized mainly along the orthorhombic *a* axis and their dispersion along the orthorhombic *c* direction could be determined. Polycrystalline samples of Nd- and Rh-doped $CeFe_2Al_{10}$ were also studied by the time-of-flight technique, with the aim of finding out how the low-energy magnetic excitation spectra change upon isoelectronic substitution of the rare-earth (Nd) on the magnetic Ce site or electron doping (Rh) on the transition-element Fe sublattice. The introduction of magnetic Nd impurities strongly modifies the spin gap in the Ce dynamic magnetic response and causes the appearance of a quasielastic signal. The crystal-field excitations of Nd, studied in both $LaFe_2Al_{10}$ and $CeFe_2Al_{10}$, also reveal a significant influence of *f*-electron hybridization (largest in the case of Ce) on the crystal-field potential. As a function of the Rh concentration, a gradual change is observed from a Kondo-insulator to a metallic Kondo-lattice response, likely reflecting the decrease in the hybridization energy.

# I. INTRODUCTION

CeFe$_2$Al$_{10}$ is a Ce based compound exhibiting Kondo insulating properties [1]. It belongs to the so-called 1-2-10 family of orthorhombic compounds Ce$M_2$Al$_{10}$ ($M$: transition metal Fe, Ru or Os) with an increase in the electrical resistivity on cooling ascribed to the opening of a narrow ''hybridization gap'' in the electronic density of states [2, 3]. All these compounds have a strongly anisotropic magnetic susceptibility. Moreover, the Ru- and Os-based systems undergo transitions to an antiferromagnetic (AF) ordered state of the rare-earth (RE) ion magnetic moments at a quite high Néel temperature $T_N$ ($T_N$ = 27 K in CeRu$_2$Al$_{10}$), while the Fe-based compound does not show any sign of magnetic ordering [1-3]. Interestingly, the direction of the ordered magnetic moment along the $c$-axis is not consistent with the easy axis in the paramagnetic state [4, 5], where $\chi_a \gg \chi_c \gg \chi_b$, corresponding to the single-ion crystal electric field (CEF) anisotropy. The value of the ordered moment is quite low with respect to that expected (according to conventional indirect exchange interaction (RKKY) and CEF models) from the $T_N$ value. These observations can be interpreted as the evidence of some competition between the electronic interactions forming the ground state of the Ce-ion and the presence of anisotropic magnetic interactions likely reflecting the anisotropic hybridization between localized 4$f$ and conduction electron states. A recent study of the magnetic excitation spectra in the ordered state (at 2 K) for CeRu$_2$Al$_{10}$ showed the existence of AF-type low-energy dispersive modes [7, 8]. Results of random phase approximation (RPA) modeling point to a discrepancy between the anisotropy of exchange parameters for these collective modes and the expectations from the single-ion anisotropy of the static susceptibility.

As was shown recently [6–10], the spin gap in the magnetic excitation spectra of the Ce-based 1-2-10 compound exists in the low-temperature limit, and gradually disappears due to formation of the quasielastic response at temperatures above 25 K, along with the disappearance of the cooperative magnetic excitation. For the classical Kondo-insulator YbB$_{12}$, it was found in (Yb,Tm)B$_{12}$ solid solutions [11] that impurity magnetic moments hinder the gap formation and suppress the in-gap excitation (resonance mode or spin-exciton[1]). In the case of SmB$_6$, another Kondo-insulator, it was shown [12] that Gd impurities on the Sm-sublattice cause an increase in the electron density of states at the Fermi energy, which can influence the gap in this compound.

In paramagnetic CeFe$_2$Al$_{10}$, a "resonance mode" (or "spin-exciton") has been reported to form at low temperature, bearing strong similarities with those occurring in the above-mentioned Kondo-insulators. In the present study, this phenomenon is studied in details in order to establish the properties of this excitation in the case of a 1-2-10 system with a nonmagnetic ground state. Polarization analysis was used in inelastic neutron scattering measurements on three-axis spectrometer (TAS) IN20 at ILL to derive the parameters of the magnetic excitation in a single crystal of CeFe$_2$Al$_{10}$.

Besides the opportunity to study the characteristics of the spin gap, the compounds of the $RE$Fe$_2$Al$_{10}$ type, with $RE$ = La, Ce, Nd, give us a chance to observe the influence of the Kondo-lattice formation on the CEF potential. Such influence has been clearly indicated for intermediate-valence compounds like CeNi [13, 14], and CeNi$_5$ [15]. It was also found recently that CEF potential is modified due to the formation of the Kondo-insulator state, as was observed for paramagnetic Tm-ion impurities substituted on the rare-earth sublattice in (Yb,Tm)B$_{12}$ [11].

---

[1] The concept of spin-exciton in the narrow gap specific for Kondo insulators was developed by Riseborough [35, 36] . This excitation mode develops due to the exchange interaction between RE-ions in Kondo-lattice.

For the present Kondo-insulator system CeFe$_2$Al$_{10}$, we set out to analyze the influence of the Nd-magnetic impurity on the gap evolution in the magnetic dynamical response function using inelastic magnetic neutron scattering. Simultaneously, we study the effect of Kondo screening on the CEF splitting of Nd ions in Ce(Nd)Fe$_2$Al$_{10}$. Interestingly, the substitution of the *d*-element in compounds from the 1-2-10 series appears to be a quite effective way of varying relevant interactions in the electronic subsystem [16, 9, 17, 18]. It was successfully applied to CeFe$_2$Al$_{10}$ in Ref. [8], where substitution of Fe for Ru was used. Rh doping is interesting in this connection [17, 18] because of the resulting change in the number of *d*-electrons. In the present work, the effect of *d*-element substitution on the Ce magnetic excitation spectra was studied on polycrystalline samples of Ce(Fe$_{1-x}$Rh$_x$)$_2$Al$_{10}$. The measurements on polycrystalline samples have been performed by the time-of-flight (TOF) neutron scattering technique for (Ce$_{0.85}$Nd$_{0.15}$)Fe$_2$Al$_{10}$ at ILL, and at ISIS with higher resolution only for pure NdFe$_2$Al$_{10}$. Samples of Ce(Fe$_{1-x}$Rh$_x$)$_2$Al$_{10}$ have been studied on the TOF spectrometer MERLIN at ISIS.

## II. EXPERIMENTAL CONDITIONS

### (a) Polarized-neutron inelastic scattering measurements on CeFe$_2$Al$_{10}$ single-crystal.

Measurements on an assembly of about 60 co-aligned single-crystals of CeFe$_2$Al$_{10}$, with total mass of about 8 g, were performed at ILL on the IN20 TAS instrument using Helmholtz coils with polarization analysis. Lattice constants of the material were $a$ = 9.01 Å, $b$ = 10.23 Å and $c$ = 9.08 Å, which is in agreement with the published data [19]. All pieces were aligned and fixed on an Al holder, $h$ = 40 mm in height and $w$ = 30 mm in width (*a* axis along *h*), on two plates with a spacing of about 10 mm (Fig. 1). We used hydrogen-free fluoropolymer CYTOP® to glue the crystals onto the Al plates. The half-width of the rocking curve on the (030) reflection was about 4 deg. In all neutron measurements, the *a* axis was perpendicular to the scattering plane (i.e. in the vertical direction). For polarization analysis, the usual convention was used, namely *x* axis parallel to the momentum transfer **Q**, *z* axis normal to the scattering plane, *y* axis completing the right-handed orthogonal coordinate system. This means that, in the case of **Q** || **b\***, i.e. the (0,k,0) direction in reciprocal space, the incident polarization **P** || **x** (denoted $P_x$) in the spin-flip channel involve scattering from the $M_y$ ($M_c$) and $M_z$ ($M_a$) components of the magnetic moments. The other two polarizations (**P** || **y** and **P** || **z**) involve the $M_a$ and $M_c$ components, respectively. All energy scans at a given momentum transfer **Q** were performed in the constant $E_f$ mode with $k_f$ = 2.662 Å$^{-1}$. Typical measuring time for one (**Q**, E) point for each polarization was about 6 minutes. The flipping ratio was about 13. An orange-type ILL cryostat was used to cool the sample down to $T_{min} \approx 2$ K.

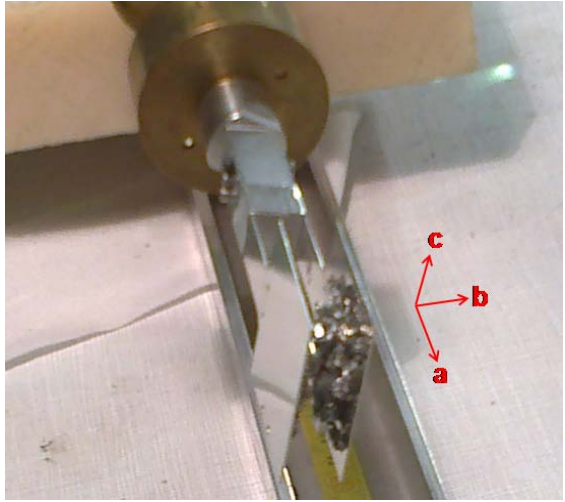

**Fig. 1.** Sample assembly and crystal orientation of $CeFe_2Al_{10}$ single crystals on the sample holder at IN20 (ILL).

Several questions were addressed in the single-crystal neutron scattering measurements using polarization analysis:

- the magnetic anisotropy of the spin excitation with energy around 10 meV existing at low temperature;
- the character of its dispersion along the (0,K,0+q) direction (not studied previously), and the $Q$ dependence of its intensity in relation to the $Ce^{3+}$ magnetic form factor;
- the temperature evolution of the parameters of the spin exciton.

*(b) Powder measurements of spin-gap and CEF effects in $(CeLa)_{1-x}Nd_xFe_2Al_{10}$*

A study of the spin-gap evolution and CEF effects in a Kondo-lattice matrix was performed on powder samples based on $LaFe_2Al_{10}$ and $CeFe_2Al_{10}$ with Nd-substitution on the Ce-sublattice. Neutron scattering measurements have been carried out on the TOF spectrometer IN4 with incoming neutron energies $E_0 = 38$ meV (resolution for elastic scattering $\Delta E_0 = 2.3$ meV) and $E_0 = 16$ meV ($\Delta E_0 = 1$ meV) at temperatures between 2 K and 80 K. With $E_0 = 16$ meV the energy resolution $\Delta E$ was better than 1 meV for the neutron energy loss part of the spectra. The measured samples were $CeFe_2Al_{10}$, $NdFe_2Al_{10}$, $Ce_{0.85}Nd_{0.15}Fe_2Al_{10}$, $La_{0.85}Nd_{0.15}Fe_2Al_{10}$, and $LaFe_2Al_{10}$ in the temperature range between 2 K and 80 K. Each temperature point took about 4 hours. The sample with 100% La was used for the evaluation of the nuclear scattering background and phonon contribution to the measured spectra. Measurements of $NdFe_2Al_{10}$ and $La_{0.85}Nd_{0.15}Fe_2Al_{10}$ provide a reference for CEF effects and their relations to the lattice spacing, approaching to the impurity limit.

For scattering angles of the detectors comprised between 13° and 118°, the momentum transfer range is approximately 0.8 Å$^{-1}$ – 5.0 Å$^{-1}$ (for $E_0 = 16$ meV), and 1.4 Å$^{-1}$ – 7.3 Å$^{-1}$ (for $E_0 = 38$ meV). Four detector groups (average scattering angles $\langle 2\Theta \rangle = 18°, 28°, 99°, 112°$) have been selected from the experimental spectra, then intensities were integrated over selected group of scattering angles (as mentioned above) after removal of the defective detector banks and of detectors contaminated by strong Bragg scattering intensity. The strongest magnetic scattering signal is expected for the first and second groups, since the magnetic scattering intensity decreases as the square of the magnetic form factor $F^2(Q)$ of Ce/Nd. The other two groups, at larger scattering angles, exhibit a higher background and a larger phonon scattering intensity as the phonon intensity vary as $\sim Q^2$.

Samples of $NdFe_2Al_{10}$ (at temperatures of 5 K and 75 K) and $LaFe_2Al_{10}$ ($T$ = 5 K) have been studied at the MARI spectrometer (ISIS) with $E_0$ = 35 meV, which provides an instrumental energy resolution of $\Delta E_0$ = 1.3 meV, significantly higher than on IN4 with a similar $E_0$.

### (c) Study of replacing Fe with Rh in $Ce(Fe_{1-x}Rh_x)_2Al_{10}$

Neutron scattering measurements have been carried out on polycrystalline samples of $RE(Fe_{1-x}Rh_x)_2Al_{10}$, (RE = La, Ce; $x$ = 0.05, 0.10, 0.20) on the MERLIN spectrometer (ISIS, RAL). The sample temperature was varied between 5 K and 150 K. Incoming neutron energies were $E_0$ = 20 and 25 meV, giving a FWHM of the elastic line of 1.3 meV and 1.7 meV respectively. Two groups of detectors were defined, hereafter denoted "low-angle" (10° < 2Θ < 25°, average  2Θ$_l$ ≈ 16°), and "high-angle" (110° < 2Θ < 135°, average  2Θ$_h$ ≈ 122°), with corresponding momentum transfers at elastic position $Q_{0l}$ = 0.9 Å$^{-1}$ and $Q_{0h}$ = 5.3 Å$^{-1}$ for $E_0$ = 20 meV, and $Q_{0l}$ = 1.0 Å$^{-1}$ and $Q_{0h}$ = 6.0 Å$^{-1}$ for $E_0$ = 25 meV.

The scattering intensity is generally assumed to be predominantly magnetic at low angle, and nuclear (phonons) at high angle. However, the above parameters indicate that, whereas "low angles" do provide good conditions for observing magnetic scattering, "high angles" are not really large enough to suppress it completely and to deliver pure phonon signal. Therefore the standard procedure for the separation of magnetic and nuclear scattering is not applicable. We will therefore make use of an alternative approach, applied earlier to the study of the relatively weak signal from the intermultiplet transition in CeNi [20] (see Section III C.2).

## III. EXPERIMENTAL RESULTS

### A. Magnetic excitations in $CeFe_2Al_{10}$ (single-crystal polarized neutron measurements)

#### 1. Magnetic anisotropy

Inelastic polarized neutron experiments have been performed on a $CeFe_2Al_{10}$ single-crystal at $T$ = 2 K for all three polarizations in the spin-flip (SF) channel, and with $P_x$ polarization only in the non-spin-flip (NSF) channel. The scattering vector studied, **Q** = (0, 3, 0) corresponds to the magnetic Brillouin zone center for AFM-ordered Ce-based 1-2-10 compounds [7]. Here it was found to provide optimal conditions for the measurement of the spin-exciton mode. The NSF-x signal, consisting mainly of nuclear scattering (see below), exhibits a monotonically increasing intensity with increasing energy transfer up to 25 meV, with some spurious counts observed near 5, 8, and 23 meV. Background correction of the SF signal at two temperatures, $T_{min}$= 2 K and 60

K was achieved in the usual way by combining data for all three incident polarizations. The obtained background was smoothed by a polynomial function prior to subtraction from the experimental spin-flip data.

The corrected spectra are presented in Fig.2, showing a pronounced anisotropy between the scattering arising from $M_a$ ($P_y$ polarization) and $M_c$ ($P_z$ polarization) magnetic moment components. Integrated magnetic intensities obtained from Gaussian fits to the spectra are in the ratio of $(M_a/M_c)^2 = 2.7 \pm 0.3$.

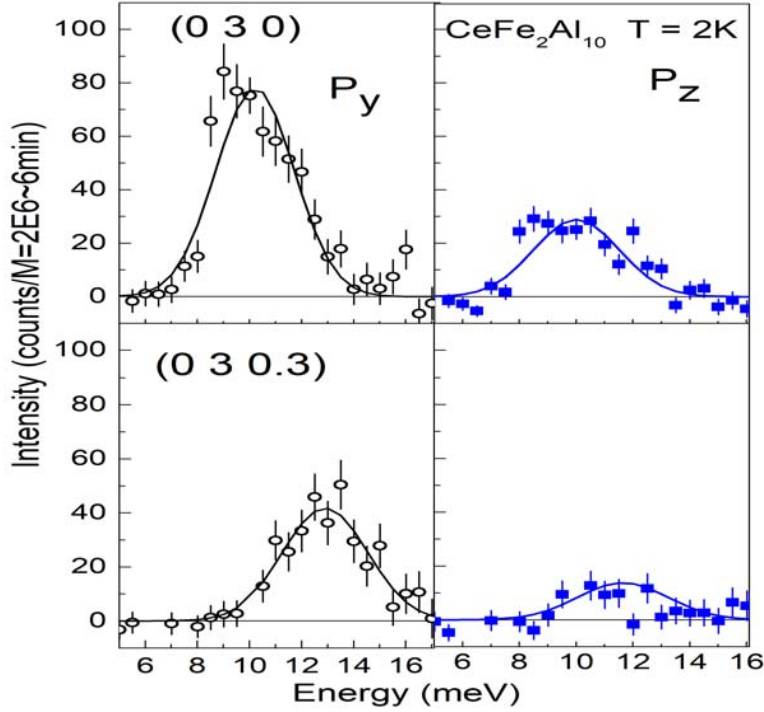

**Fig.2.** Background-corrected INS spectra for single-crystal CeFe$_2$Al$_{10}$ at $T = 2$ K, measured in the spin-flip channel for incident neutron polarizations $P_y$ and $P_z$ at **Q** = (0, 3, 0) and (0, 3, 0.3). Lines represent Gaussian fits to the data.

These results can be compared with the single-crystal static magnetic susceptibility data [8,19, 21]. The temperature dependence $\chi(T)$ displays significant anisotropy, thought to originate mainly from single-ion crystal-field effects [22, 23], which develop as temperature decreases. The similarity between neutron and static susceptibility data thus suggests that the anisotropy of the $q$-dependent excitation spectra in the spin-gap regime at low temperature may also result from crystal-field effects. Excitations also show clear dispersion over the Brillouin zone as discussed below.

## 2. Dispersion of the spin-exciton

With the present orientation of the sample, the dispersion of the spin-exciton branch could be studied along the $c^*$ direction, which had not been done previously. Experimental data for two reduced **q** vectors are presented in Fig. 2. The use of polarized neutrons provides optimal conditions for separating the relatively weak magnetic signal from the phonon background which is almost totally suppressed in the SF channel. Fig. 3 presents the **Q** dependence of the excitation energy along the $(0, 3, q)$ direction, as derived from the Gaussian fit to the (background-corrected) experimental spectra. For the selected direction, $P_x$, of the incident neutron polarization, the neutron scattering intensity results from fluctuations of the $M_a$ and $M_c$ components of the Ce magnetic moments.

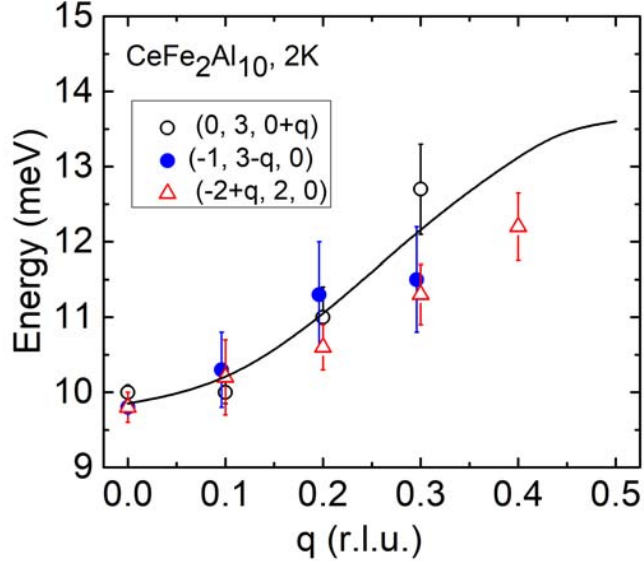

**Fig. 3.** Energy dispersion of the spin-exciton mode along different reciprocal space directions of CeFe$_2$Al$_{10}$, derived from the magnetic neutron scattering spectra at 2 K. Open circles denote data along the $c^*$ direction $(0, 3, q)$ from the present experiments (SF$_x$ channel). Also shown are the dispersions along the $a^*$ (open triangles) and $b^*$ (closed circles) directions obtained in unpolarized TAS at LLB [10]. The line is a guide to the eye as in Ref. [10].

In Fig. 3, the present data for the $c^*$ direction are compared with those obtained along $a^*$ and $b^*$ in a previous unpolarized neutron experiment [10], performed on a different CeFe$_2$Al$_{10}$ single-crystal. One sees that the amplitude and shape of the dispersions are comparable. Interestingly, the dispersion also looks qualitatively similar to that observed for the in-gap magnetic excitation of CeRu$_2$Al$_{10}$ in the AFM ordered state [7]. It can be noted that the intensity of the magnetic signal is strongly suppressed for $q > 0.3$ r.l.u. The maximum energy of the excitation near the magnetic zone boundary is in agreement with the peak position, around 14 meV, in the powder spectra of CeFe$_2$Al$_{10}$ [9], corresponding to the modes near the high-energy edge contributing with a large spectral weight.

*3. Temperature dependence*

It is well documented that resonance or spin-exciton modes observed experimentally in the low-temperature magnetic excitation spectra of various Kondo-insulators [24, 25, 26] are dramatically suppressed by an increase of the temperature. The complete damping of the mode typically takes place at temperatures much lower than its excitation energy. The temperature dependences of the energy and intensity of the spin-exciton in $CeFe_2Al_{10}$ determined in the present experiments are presented in Fig. 4.

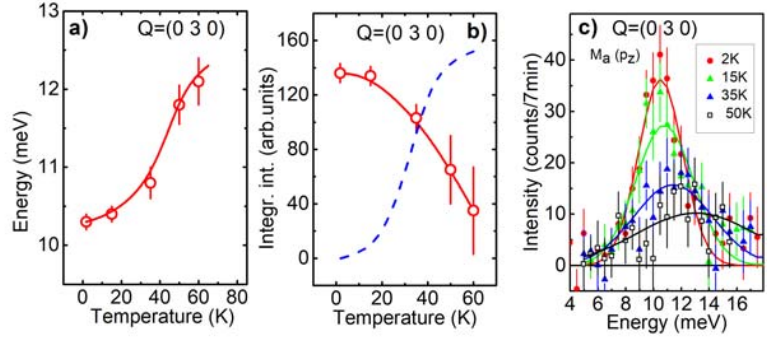

**Fig. 4.** Temperature dependence of (a) the energy and (b) the intensity of the spin-exciton mode ($M_a$ component) in $CeFe_2Al_{10}$ derived from the intensity measured for **Q** = (0, 3,0) in the $SF_y$ channel. Lines are guides to the eye. The dashed line in (b) is a smoothed approximation of the data for the quasielastic peak intensity from Figs. 7 and 11 (see below). (c) Background-corrected INS spectra at various temperature measured with $Pz$ incident neutron polarization at **Q** = (0, 3, 0). Lines represent Gaussian fits to the data.

With increasing temperature, the energy of the mode at the magnetic zone center plotted in frame (a) increases, causing a reduction of the total amplitude of the dispersion. Meanwhile, the intensity of the peak is steeply suppressed [frame (b-c)]. Measurements of the quasielastic signal in $CeFe_2Al_{10}$ powder, realized on IN4 and MERLIN, respectively, are presented in details in sections B2 and C1 below. In Fig. 4(b), an approximation of the temperature dependence of the quasielastic intensity is shown as a dashed line for comparison with that of the inelastic signal. The mirror-like behaviors of those two quantities is typical for Kondo insulators.

In the present measurements, we failed to observe any significant quasielastic signal at high temperature in either the SF or the NSF channel, contrary to the earlier unpolarized neutron measurements at LLB [10] as well as that observed [9] for powder samples of $CeFe_2Al_{10}$. The reason for this discrepancy may be that, because the magnetic fluctuations involved are widely spread out in **Q** space, the scattered intensity becomes undetectable using the comparatively weak polarized neutron beam.

**B. Effect of Nd substitution (TOF powder measurements on IN4)**

*1. Separation of the magnetic scattering signal*

To analyze the quasielastic magnetic signal from the Ce ions, we mainly focused on data collected in the first detector groups of IN4, with an average scattering angle $2\Theta = 18°$, using an incoming neutron energy $E_0 = 16$ meV (energy resolution $\Delta E$ of less than 1 meV on the neutron energy loss side). The phonon signal at low energy transfer is expected to be about three times stronger in the La-based sample because it arises primarily from the soft rare-earth vibration modes, and the ratio of La- to Ce- neutron scattering cross sections is close to 3.

In Fig. 5, the spectral intensities for $LaFe_2Al_{10}$ and $CeFe_2Al_{10}$ are shown for the lower-angle detector group at two temperatures, $T = 2$ K and 60 K. The temperature increase in the intensity associated with the phonon density of states below 10 meV can be estimated from the comparison of the $LaFe_2Al_{10}$ spectra [Fig. 5(a)]. The difference remains within the limits of error bars for positive energy transfers, while some difference exists at negative energies. This result indicates that the phonon contribution to the scattering intensity at positive energy transfer up to $E \approx 10$ meV is very weak for the scattering angle range under consideration.

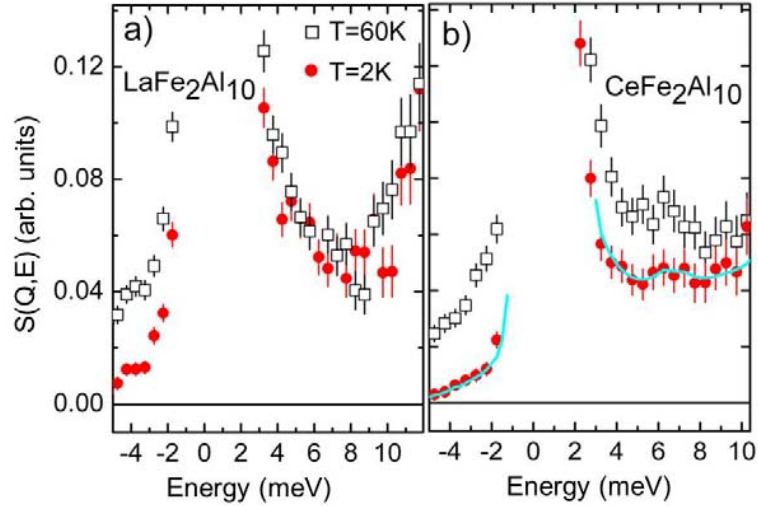

**Fig. 5.** Spectral function $S(\mathbf{Q}, E)$ for (a) $LaFe_2Al_{10}$ and (b) $CeFe_2Al_{10}$ at $T = 2$ K (circles) and 60 K (squares), measured with incoming energy $E_0 = 16$ meV at $2\Theta = 18°$ average scattering angle. The cyan line in frame (b) represents a smoothing of the $CeFe_2Al_{10}$ data measured at 2 K.

The extra intensity in the foot of elastic line observed in both compounds between 3 and 6 meV most likely originates from double elastic neutron scattering on the sample environment, or other similar elastic processes. The existence of such contamination in inelastic spectra at low positive energy transfers is a well-known effect in neutron TOF experiments [27]. This contribution has to be subtracted out before a quantitative analysis of the Ce-spectral function can be performed. In $CeFe_2Al_{10}$ [see Fig.5(b)], the additional increase in the intensity below 8 meV occurring when temperature is raised from 2 to 60 K cannot be due to phonon scattering, but rather reflects a temperature increase in the quasielastic magnetic response [9, 10].

Previous experiments [10] have established the absence of quasielastic magnetic scattering in $CeFe_2Al_{10}$ at low temperatures, at least up to 20 K. Since the direct phonon contribution at this temperature is considered to be negligible, as discussed above, the spectrum measured at $T = 2$ K, at least up to the lower edge of the dispersive mode near 8 meV, can be taken to represent the temperature-independent background for the given sample-spectrometer configuration. That spectrum was smoothed, as shown by the solid line in Fig. 5(b), normalized according to the ratio of sample masses in the case of $Ce_{0.85}Nd_{0.15}Fe_2Al_{10}$, then finally subtracted from the measured data for $CeFe_2Al_{10}$ and Nd-substituted samples. It can be noted that this smoothed function is quite similar to the 2 K spectrum for the nonmagnetic $LaFe_2Al_{10}$ sample, within a numerical factor essentially reflecting differences in sample masses and monitor counts.

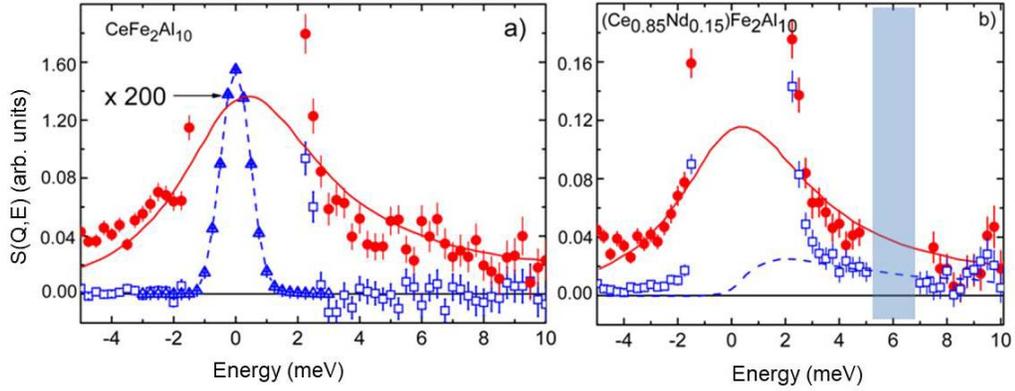

**Fig. 6.** Magnetic scattering function for $CeFe_2Al_{10}$ (a) and $Ce_{0.85}Nd_{0.15}Fe_2Al_{10}$ (b) (blue open symbols: $T = 2$ K, red closed symbols: $T = 60$ K). Triangles in (a) represent the elastic line on a reduced intensity scale. Lines are quasielastic Lorentzian fits to the data, with temperature factor included [see eqn. (1) below]. The hatched area around 6 meV represents the position of $Nd^{3+}$ CEF peak, which was excluded from the fit.

*2. Quasielastic magnetic response*

The resulting spectra displayed in Fig. 6 represent our best estimate of magnetic scattering in $CeFe_2Al_{10}$ and $Ce_{0.85}Nd_{0.15}Fe_2Al_{10}$ compounds at $T = 2$ and 60 K. While they may still include some weak residual component from phonon scattering at the higher temperature, this effect should be essentially negligible. As implied by the subtraction procedure used, the 2 K spectrum for $CeFe_2Al_{10}$ reduces to the scatter of experimental data points around the smoothed line.

In the case of $CeFe_2Al_{10}$ [Fig. 6(a)] a significant temperature effect is observed. We estimate that 90 to 95% of this effect results from the appearance of quasielastic magnetic scattering as temperature increases. The data measured at energy transfers comprised between -5 and +8 meV can be used for fitting the quasielastic response, only excluding the interval from -2 to +3 meV, which is contaminated by the elastic line.

Lorentzian fits, using eqn. (1), to the data measured at 2 and 60 K are shown as lines in Fig. 6. The energy dependence of the quasielastic neutron scattering intensity $I$ at temperature $T$ is given by

$$I(E) = A_q(\Gamma_q/2) \, E \, [(\Gamma_q/2)^2 + E^2]^{-1} / [1-\exp(-E/k_BT)], \qquad (1)$$

where $A_q$ is an amplitude and $\Gamma_q$ is the line width (FWHM) of the Lorentzian component centered at $E = 0$ meV. At $T = 60$ K, the residual phonon contribution slightly reduces the quality of the fit for negative energy transfers.

Figure 7 shows the temperature dependences of $\Gamma_q$ (left) and of the integrated intensity $S_q$ (right), defined as the area under the Lorentzian spectral function $S_q = 1.57 \, A_q \Gamma_q$ for all measuring temperatures. The temperature evolution of the quasielastic magnetic signal looks qualitatively different in the pure and Nd-substituted compounds. This difference can be ascribed to the effect of Nd magnetism on the Ce moment fluctuations, since the scattering signal from Nd ions themselves is expected to be quite narrow, likely resolution limited, in view of the width of CEF peaks (see below).

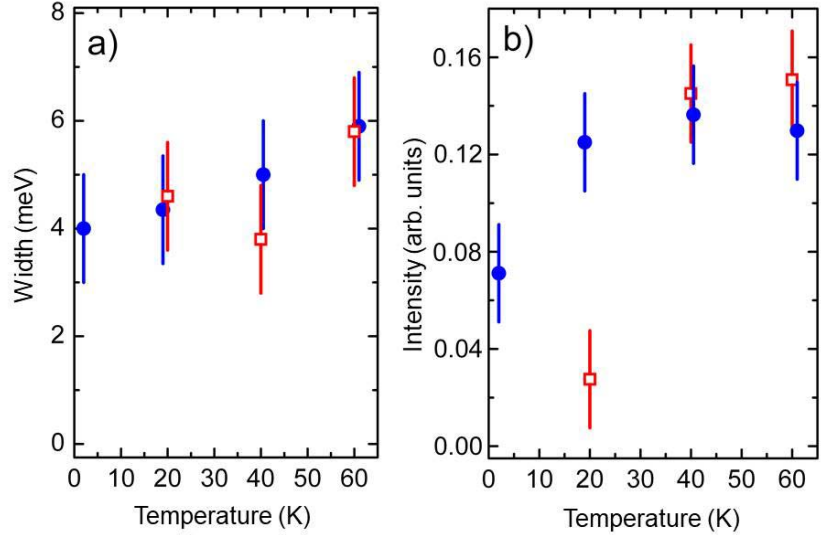

**Fig. 7.** Temperature dependences of the linewidth, FWHM (a) and integrated intensity (b) of the quasielastic peak in $CeFe_2Al_{10}$ (open squares) and $(Ce_{0.85}Nd_{0.15})Fe_2Al_{10}$ (closed circles).

In $CeFe_2Al_{10}$, the temperature enhancement of the quasielastic intensity measured on IN4 [Fig. 7(b)] mirrors the suppression of the spin exciton previously observed in the IN20 measurements [Fig. 4(b), above]. In the sample containing Nd, the quasielastic signal from Ce exists even at the lowest temperature ($T = 2$ K) and its increase to a saturation value takes place in a narrower temperature interval (up to about 20 K) than in the pure compound (saturation around 40 K).

### 3. Influence of Kondo screening on the CEF splitting of Nd ions in $Ce(Nd)Fe_2Al_{10}$

In this section, we focus on the low-temperature TOF data for $(La,Ce,Nd)Fe_2Al_{10}$ compounds obtained on MARI (ISIS) and IN4 (ILL).

The spectral function for NdFe$_2$Al$_{10}$ measured on MARI at temperatures $T$ = 5 K and 75 K is shown in Fig. 8. Only data for the lower scattering angle group (momentum transfer $Q$ < 3A$^{-1}$) have been plotted. Four inelastic peaks are clearly seen in the low-temperature spectrum, corresponding to transitions from the ground state to four doublet excited states. This pattern reflects the maximum splitting of the Nd$^{3+}$ $J$ = 9/2 ground state multiplet into 5 CEF doublets allowed by Kramers theorem in a low-symmetry CEF potential with the orthorhombic point group, C$_{2v}$. The energies of the doublets observed in the measurement are 0 – 5.9 – 11.5 – 16.1 – 17.7 meV, with an average accuracy of about 0.1 meV. The total energy splitting is consistent with static magnetic measurements [22], and lies in the energy range of 20 - 30 meV typical for most Nd-based intermetallic compounds. The two peaks at higher energies could be clearly separated only thanks to the high energy resolution of the instrument.

Increasing temperature causes additional peaks to appear, some of which are located quite close to each other, or to those already present at $T_{min}$ = 5 K. In order to assign each high-temperature peak to the correct CEF transition, one needs to make use of the information already derived from the peak structure at $T_{min}$.

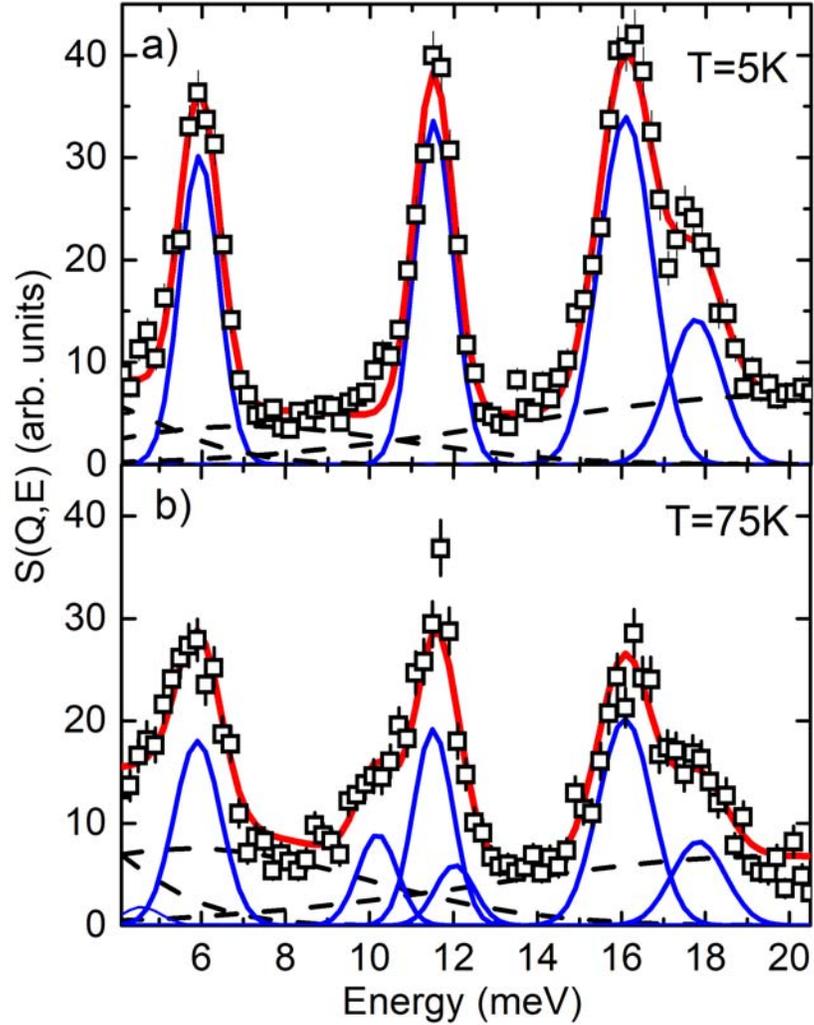

**Fig. 8.** Neutron spectral functions of NdFe$_2$Al$_{10}$ at $T = 5$ K (a) and 75 K (b), measured on MARI (ISIS) with incident energy $E_0 = 35$ meV. Lines represent fits using Gaussian spectral components. The energies of the peaks at $T = 2$ K denote the positions of all four excited doublets, split from the $J = 9/2$ ground state multiplet of Nd$^{3+}$ by the orthorhombic CEF potential. Three broad peaks (dotted lines) represent the phonon contribution, as estimated from the measurement of LaFe$_2$Al$_{10}$.

The results of the measurements of NdFe$_2$Al$_{10}$, La(Nd)Fe$_2$Al$_{10}$, and Ce(Nd)Fe$_2$Al$_{10}$ on IN4, with incident energy $E_o = 38$ meV, are shown in Fig. 9(a). Data are plotted for the group of detectors at the lowest scattering angles ($13° < 2\Theta < 23°$), after normalization to the sample mass, correction for neutron transmission, and phonon subtraction. The phonon scattering contribution had to be derived from the LaFe$_2$Al$_{10}$ spectra since the momentum transfer range accessible in the present measurement transfer (less than 8.5 Å$^{-1}$ for the highest scattering angles $2\Theta = 111°$ at energy transfer $E = 20$ meV) was not sufficient to ensure pure nuclear scattering in the case of the Nd-based sample. The procedure is based on the assumption that the low-energy peak (at about 11 meV) in the phonon spectrum of $RE$Fe$_2$Al$_{10}$ mainly arises from vibration modes of the rare-earth ion, whereas other peaks at higher energies (up to 30 meV) reflect the phonon density of states due to the (Fe$_2$Al$_{10}$) lattice. An approximate spectral function for phonon scattering in a given 1-2-10 compound can thus be obtained by scaling the intensity of the lower peak according to the ratio of the total neutron nuclear scattering cross-section of the rare-earth. This reconstructed phonon spectrum was then normalized to the experimental data in the energy range $E > 20$ meV, where only phonon contribution exists, and subtracted from the measured intensities to obtain the magnetic spectral function. In Fig. 9(b), the final results have been normalized to the Nd content in each sample.

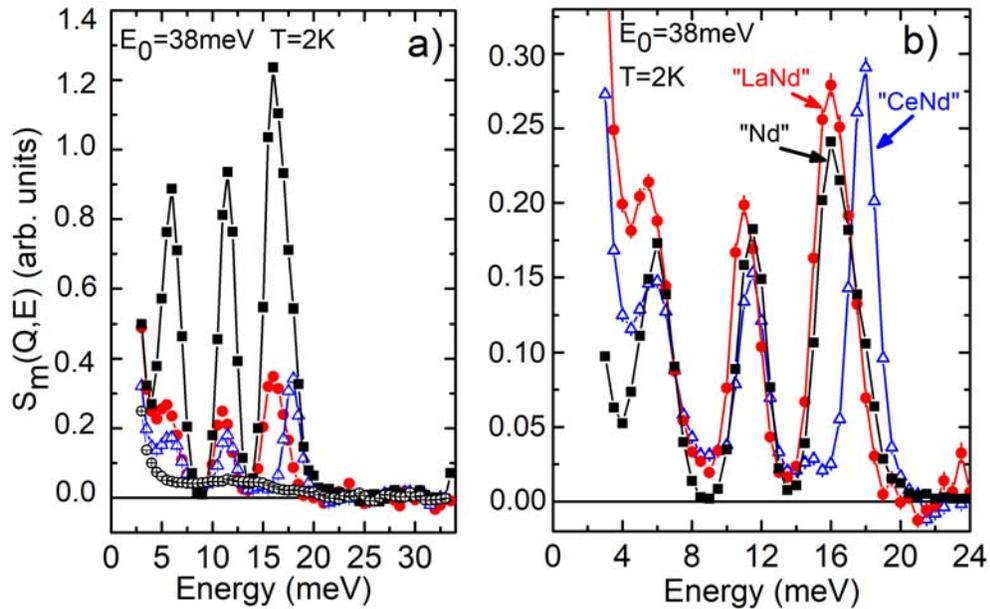

**Fig. 9.** Magnetic spectral functions for NdFe$_2$Al$_{10}$ (black squares), La$_{0.85}$Nd$_{0.15}$Fe$_2$Al$_{10}$ (red circles), Ce$_{0.85}$Nd$_{0.15}$Fe$_2$Al$_{10}$ (blue triangles), and CeFe$_2$Al$_{10}$ (open circles) at $T = 2$ K, measured on IN4 with incident neutron energy $E_0 = 38$ meV. Experimental spectra in frame (a) are normalized to the sample mass; in frame (b) they are further normalized to the Nd concentration in the solid solution.

In Fig. 10, low-angle ($13° < 2\Theta < 23°$) spectra, measured a lower incoming energy of 16 meV, are presented for two different compounds: La$_{0.85}$Nd$_{0.15}$Fe$_2$Al$_{10}$ and Ce$_{0.85}$Nd$_{0.15}$Fe$_2$Al$_{10}$. For energy transfers $E \leq 14$ meV, two CEF peaks from Nd$^{3+}$ are observed. Also shown is the spectrum for pure LaFe$_2$Al$_{10}$, emphasizing that no detectable phonon scattering exists under the present conditions. With the improved resolution, a clear shift in energy between the (LaNd) and (CeNd) compounds is observed for both low-energy CEF peaks, confirming the effect already visible on the higher energy (14 – 20 meV) peaks in Fig. 9.

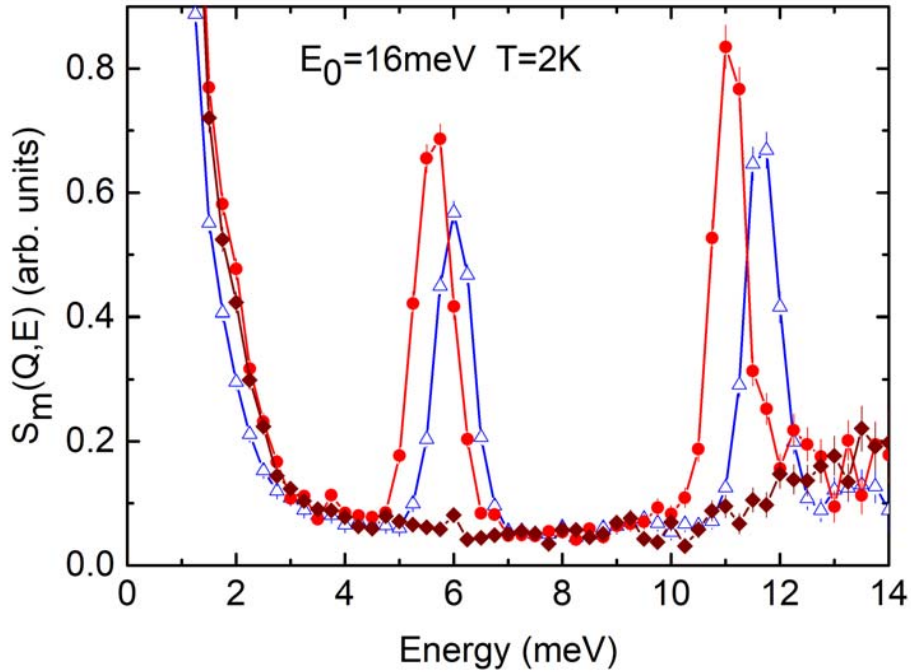

**Fig. 10.** Spectral functions for La$_{0.85}$Nd$_{0.15}$Fe$_2$Al$_{10}$ (red closed circles), Ce$_{0.85}$Nd$_{0.15}$Fe$_2$Al$_{10}$ (blue open triangles), and LaFe$_2$Al$_{10}$ (brown closed diamonds) at $T = 2$ K, measured on IN4 at low scattering angles ($13° < 2\Theta < 23°$) with incoming energy $E_0 = 16$ meV. Instrumental background has been subtracted out.

If one compares the three compounds studied, the most pronounced difference occurs between Nd ions substituted in either a Ce- or La-based matrix. This is best evidenced by the pronounced shift to higher energies of the 16 meV peak in Ce$_{0.85}$Nd$_{0.15}$Fe$_2$Al$_{10}$ as compared to La$_{0.85}$Nd$_{0.15}$Fe$_2$Al$_{10}$ [Fig. 9(b)]. This is particularly remarkable since these compounds are the ones with the closest lattice constants. On the other hand, only a minor change is found between the latter compound and pure NdFe$_2$Al$_{10}$, which can be traced back to the difference in lattice constants. To get a quantitative estimate of the effect for all measured CEF transitions, we have fitted the experimental spectra using Gaussian line shapes. The resulting energies are collected in Table I.

**Table I.** Energies ($E_i$) of inelastic CEF peaks in pure NdFe$_2$Al$_{10}$, and deviations ($\Delta E_i$) from these values for the corresponding energies in the (La,Nd) and (Ce,Nd) solid solutions at $T = 2$ K.

| $i$ | $E_i$ [NdFe$_2$Al$_{10}$] (meV) | $\Delta E_i$ [La$_{0.85}$Nd$_{0.15}$Fe$_2$Al$_{10}$] (meV) | $\Delta E_i$ [Ce$_{0.85}$Nd$_{0.15}$Fe$_2$Al$_{10}$] (meV) |
|---|---|---|---|
| 1 | 5.9 | -0.1 | +0.2 |
| 2 | 11.5 | -0.4 | +0.1 |
| 3 | 16.1 | -0.2 | **+1.9** |
| 4 | 17.7 | -0.5 | +0.3 |

The magnitude of the peak energy shifts expected from the difference in ionic radius between the RE elements can be roughly estimated. For both CeFe$_2$Al$_{10}$ and LaFe$_2$Al$_{10}$, the lattice constant *increases* with respect to NdFe$_2$Al$_{10}$, but the effect is only 0.1 to 0.2 % for all three orthorhombic directions [19, 28, 29]. Therefore, the resulting shift in the energies of the CEF transitions, caused by the change in the CEF parameters $B_{nm}$ ($\Delta B_{nm}/B_{nm} \sim (n+1) \Delta R/R$, where $R$ is the interatomic distance), should not exceed 0.3 meV for the highest level $E_4$. This estimation results from the small difference in lattice spacings (order of 0.1%) and averaged contribution of all $B_{nm}$ (effective n~4) into the value of CEF splitting. Furthermore, the energies should be *lower* than in NdFe$_2$Al$_{10}$ because the Coulomb potential is reduced when interatomic distances increase. The difference in lattice spacing between the three compounds under consideration implies that the energy of the CEF peaks should be reduced (by no more than 0.3 meV) in Ce(La,Nd)Fe$_2$Al$_{10}$ with respect to NdFe$_2$Al$_{10}$. This is indeed observed for LaFe$_2$Al$_{10}$ (Table I), but the effect is reversed in CeFe$_2$Al$_{10}$. Therefore, the enhanced CEF splitting of Nd$^{3+}$ ion in the Ce-based compound is clearly anomalous. We argue that it is the fingerprint of Kondo-lattice effects which are discussed in Section IV.1.

## C. Effect of Rh-substitution in Ce(Fe$_{1-x}$Rh$_x$)$_2$Al$_{10}$

### 1. Magnetic quasielastic spectra of Ce(Fe$_{1-x}$Rh$_x$)$_2$Al$_{10}$ at T = 5 – 100 K

In the IN4 experiments presented above, the phonon spectra at energies below 15 meV exhibit a rather weak feature, with a maximum at about 11 meV, mainly resulting from rare-earth-ion vibrations. The next peak in the phonon density of states is located near 17-18 meV, and can be ascribed to *d*-metal vibrations. The data obtained using the MERLIN spectrometer at large enough momentum transfer (see $2\Theta_h$ values in Section II) for Ce(Fe$_{1-x}$Rh$_x$)$_2$Al$_{10}$ and La(Fe$_{0.8}$Rh$_{0.2}$)$_2$Al$_{10}$ samples indicate a quite weak influence of Rh substitution on the latter phonon component. The effect is limited to a moderate increase in the peak width, along with a decrease in its intensity due to the about 2.5 times smaller neutron scattering cross-section of Rh as compared to Fe.

The temperature dependent magnetic contribution to the neutron scattering spectra of CeFe$_2$Al$_{10}$ obtained on MERLIN is shown in Fig. 11 for different sample temperatures. For $T = 5$ K the data below 9 meV fluctuate around zero intensity, a trivial consequence of using the smoothed spectrum at this temperature as an estimate of the background (same procedure as in section III.B.1 for the treatment of IN4 data). The appearance of a quasielastic signal in CeFe$_2$Al$_{10}$ can be seen at 40 K as shown in Fig. 11, which is in agreement with Fig. 4(b). At higher temperatures the full energy transfer range is covered by a rather large quasielastic signal. A fit of the data using quasielastic Lorentzian spectral functions was obtained by adjusting the amplitudes while setting the linewidths to the values derived from the IN4 experiments [Fig. 7(a)].

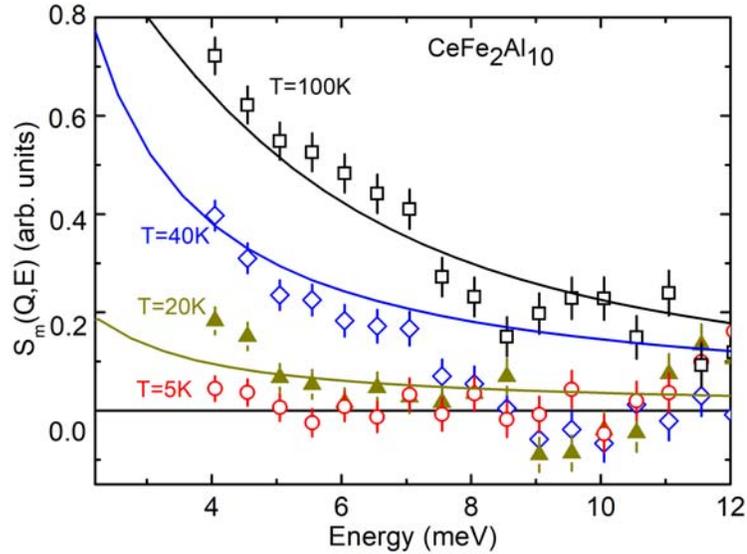

**Fig. 11.** Magnetic scattering spectra ($E_0 = 20$ meV) for CeFe$_2$Al$_{10}$ measured on MERLIN at low scattering angles $2\Theta_l$ in the temperature range 5 – 100 K, obtained after background subtraction. The solid lines represent the results of fits using quasielastic Lorentzian spectral functions, whose (fixed) line widths were derived from the treatment of IN4 data [Fig. 7(a)]. It is to be noted that the spectral functions are centered a zero energy and only positive energy transfers are shown.

One sees that the line widths determined on IN4 are in good agreement with the MERLIN data for this sample. We therefore believe that the same procedure can be reliably applied to the Rh-substituted Ce(Fe$_{1-x}$Rh$_x$)$_2$Al$_{10}$ compounds as well. The magnetic spectra of the samples with Rh concentrations $x = 0.05$, 0.10, and 0.20, after background subtraction, are shown in the Fig.

12. The solid lines represent the results of quasielastic Lorentzian fits, whose parameters (line widths and integrated intensities) are plotted in Fig. 13 as a function of the Rh content.

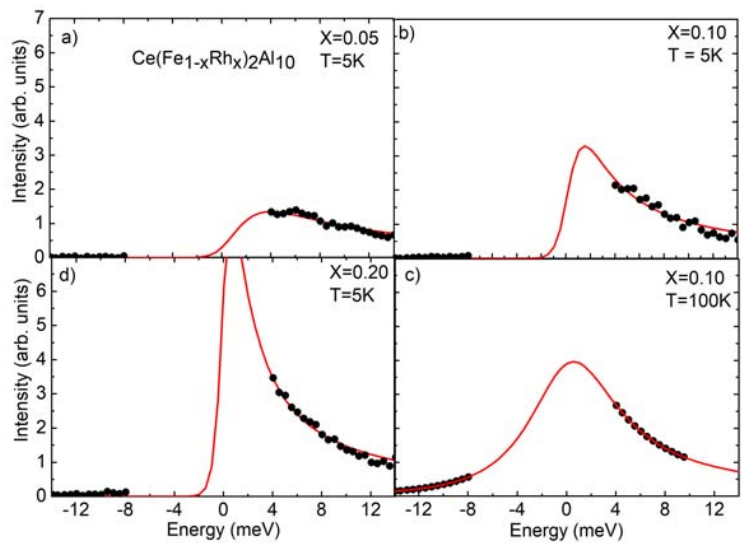

**Fig. 12.** Magnetic excitation spectra of the Ce(Fe$_{1-x}$Rh$_x$)$_2$Al$_{10}$ samples for $x$ = 0.05, 0.10, and 0.20 at $T$ = 5 K and (for x = 0.1 only) $T$ = 100 K, after background subtraction (closed circles). The lines represent fits by quasielastic Lorentzian spectral functions.

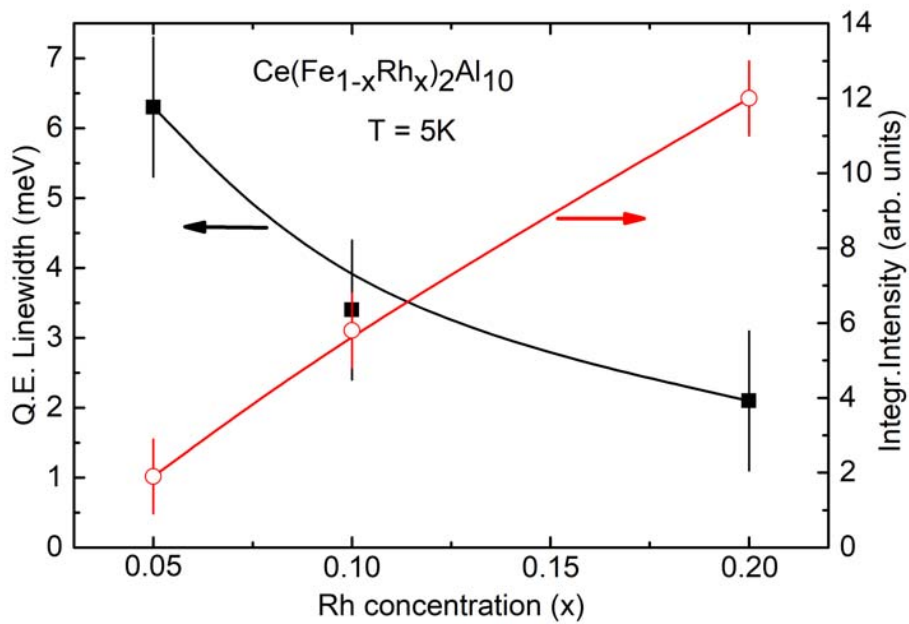

**Fig. 13.** Fitted parameters of the quasielastic magnetic spectral component at $T = 5$ K (line width, FWHM: closed squares; integrated intensity: open circles) as function of the Rh-concentration $x$ in $Ce(Fe_{1-x}Rh_x)_2Al_{10}$.

It is seen from Fig. 13 that increasing the Rh concentration results in the appearance and increase of a quasielastic signal at the lowest measured temperature of 5 K and the simultaneous decrease of the line width. For $x = 0.05$, the line width is comparable to that of the quasielastic line in pure $CeFe_2Al_{10}$ above 60 K, (6±1 meV), reported above in Section III.B.2. We recall that, in the latter compound, no quasielastic signal exists below 20 K because of the spin-gap formation (Fig. 11).

*2. Inelastic response in $Ce(Fe_{1-x}Rh_x)_2Al_{10}$ at $T = 5 – 100$ K*

In the present section, we seek evidence for an inelastic contribution in the Rh-doped compounds $Ce(Fe_{1-x}Rh_x)_2Al_{10}$. For $x = 0$ (pure $CeFe_2Al_{10}$), it was shown from single-crystal measurements (Ref. [10] and Section III.A of this paper) that a branch of dispersive excitations exists in the energy range 10-15 meV. In the powder measurements, a maximum in the intensity of the inelastic signal was experimentally observed at about 14 meV [9].

In the present powder spectra of $Ce(Fe_{1-x}Rh_x)_2Al_{10}$ at $T = 5 – 100$ K measured on MERLIN with $E_0 = 20$ or 25 meV, there is no obvious evidence of an inelastic peak even at the lowest concentration $x$. The reason for this may be both physical (a relatively broad energy width of the inelastic spectral component averaged in **Q** space) and experimental, e.g. the low incoming neutron energy, which restricts the accessible $Q$-$E$ space and results in poor signal-to-background ratio, and/or the energy-dependent background.

To extract experimental information on the inelastic magnetic signal, we used a difference method by combining spectra measured at different temperatures, in analogy with the data treatment performed an earlier work on intermultiplet excitations [20]. The idea is that, for $CeFe_2Al_{10}$, the maximum in intensity of the magnetic inelastic signal occurs at the lowest temperatures, below 20 K [9, 10]. Therefore, if one subtracts the 5 K data from those measured at higher temperatures, negative values in the difference spectrum over some energy range could denote an extra inelastic signal developing at low temperature.

Such a feature is actually seen on the MERLIN data for $CeFe_2Al_{10}$, despite the quite low incoming neutron energy ($E_0 = 20$ meV) with respect to the excitation energy of 12 meV [Figs. 11 and 14(a)]. Systematic negative values in the difference spectra $S_d(E, T)$ are observed at $T = 40$ K around 12 meV energy transfer. This temperature is just in the temperature region where the inelastic signal intensity is mainly suppressed, as shown in Fig. 4. With further increase of the temperature, this dip is wiped out owing to the increasing contribution from magnetic quasielastic and phonon scattering (for $T$ above 60 K).

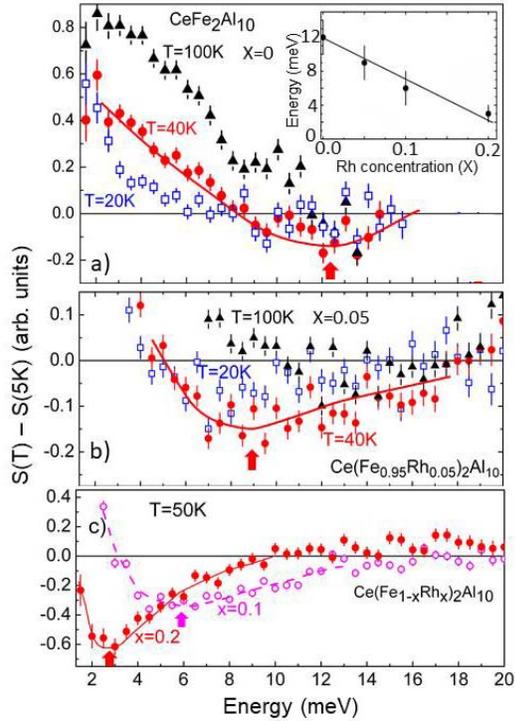

**Fig. 14.** Difference spectra, $\Delta S(T) = S(T) - S(5K)$, for $Ce(Fe_{1-x}Rh_x)_2Al_{10}$ samples with x = 0 measured with E0 = 20 meV (a), x = 0.05 (b), 0.1 and 0.2 (c), measured with E0 = 25 meV at "low scattering angle" ($10° < 2\Theta < 25°$, $\langle 2\Theta_l \rangle \approx 16°$). Frames (a) and (b): temperatures are T = 20 K (open squares), T = 40 or 50 K (closed circles), and T = 100 K (closed triangles). In panel (c), data for x = 0.1 (open circles) and x = 0.2 (closed circles) are displayed for one temperature: T = 50 K. Lines are guides to the eye. Arrows indicate the energy positions of the minima. Inset in (a): inelastic peak energy as a function of Rh concentration at T = 5 K.

In the case of the Rh-doped compounds ($x$ = 0.05, 0.1, and 0.2), the same procedure was applied to the data collected in the low-scattering-angle groups of detectors. Again, clear evidence for an inelastic spectral component in seen in the panels (b) and (c) of Fig. 14. Increasing the Rh concentration gradually shifts the position of the negative minimum occurring around 40 – 50 K in the difference spectra, which roughly corresponds to the energy of the maximum of the inelastic peak at $T$ = 5 K. This peak is located at 9 ± 2 meV, 6± 2 meV, and 3± 1 meV for $x$ = 0.05, 0.1, and 0.2, respectively. This concentration dependence, plotted in the inset of Fig. 14(a), is seen to be quite pronounced.

## IV. DISCUSSION

In the above measurements, substitutions of a rare-earth element (Nd) for Ce, and of a transition element (Rh) for Fe have been used to gain insight into the spin dynamics of the Kondo insulator state in $CeFe_2Al_{10}$. In both cases, the low-temperature spin gap was found to be suppressed while a quasielastic component appears. However, other features differ significantly. In the case of Nd substitution, CEF transitions dominate the low-energy magnetic excitation spectra, making it impossible to trace the evolution of the spin-exciton mode. However, a careful analysis of these excitations reveals interesting effects, which reflect the hybridization of the Ce 4*f* orbitals. In the case of Rh substitution, on the other hand, a weak inelastic signal reminiscent of the spin-exciton mode could be detected, whose evolution points to a shift of this excitation to lower energies with increasing Rh concentration.

**1.** *Nd CEF as a sensor of hybridization phenomena in 1-2-10*

In rare-earth based intermetallic compounds, the so-called "CEF effects" actually involve several physical mechanisms such as Coulomb interactions or covalence. As a result, their strength depends on multiple factors. The most obvious one is the distance to the neighbors on the crystal lattice. The influence of the lattice spacing on the CEF parameters had been characterized in early neutron spectroscopy studies of CEF splitting in metals, for instance in the $Pr_{1-x}(La,Y)_xAl_3$ series [30, 31]. The general trend of this effect appears to be in accordance with the variation of the Coulomb potential produced by changes in lattice spacing (in the latter work, the range of lattice spacing variation was up to 3%, more than 10 times that in the present system), namely an increase in the CEF splitting with decreasing interatomic distances.

In addition, very strong hybridization effects have been observed in the Ce-based intermediate-valence compounds $CeNi_5$ [15, 32] and CeNi [14], using magnetic $Pr^{3+}$ and $Nd^{3+}$ ions, respectively, as sensors of the CEF potential. In (Ce,Nd)Ni, the influence of intermediate valence Ce-ions on the CEF potential manifests itself in a *reduction* of the $Nd^{3+}$ CEF splitting, opposite to the effect expected from the *decrease* in lattice spacing with respect to NdNi. The effect is even more dramatic in the strongly intermediate-valence compound $CeNi_5$. As shown in Ref. [32], the strongest effect occurs in the $\Gamma_3 - \Gamma_4$ splitting of the $Pr^{3+}$ ground-state multiplet, which is defined (because of local symmetry) by the $B_{66}$ term in the CEF Hamiltonian. This term corresponds to the $l = 3$ (i.e. *f*-type) electron contribution to the CEF potential. The substitution of Ce for Pr in $(Ce,Pr)Ni_5$ results in a decrease by almost 30% of the $\Gamma_3 - \Gamma_4$ CEF excitation energy of $Pr^{3+}$ ion, superimposed on the normal 8% increase expected due to the reduction of the lattice spacing. It has been established experimentally that, in the integral-valence structural analogue $(Pr,Y)Ni_5$, the lattice contraction produces a trivial increase in the CEF splitting.

The above observations, made on intermetallics, provide a useful background to understand the present results for Ce 1-2-10 (as well as for $YbB_{12}$ [11]) Kondo insulators. For instance, the strong renormalization of the $E_3$ energy (see Table 1), along with those of other CEF transitions for $Nd^{3+}$ in $CeFe_2Al_{10}$, opposite to the normal effect of lattice spacing dominated by the change in the Coulomb potential, as observed for $Nd^{3+}$ in $LaFe_2Al_{10}$, can be ascribed to the peculiar *f*-electron hybridization phenomena taking place in Ce-based compounds. This phenomenon is thought to influence the band-electron contribution to the CEF potential. Its effect is particularly strong in the above mentioned intermediate-valence systems, but a similar, though less pronounced, influence is also expected to occur in Kondo-insulator systems with the relatively low extent of f-electron instability or weaker 4f-conduction electron hybridization.

CEF studies in Kondo insulators in general are scarce. The effects of hybridization were recently investigated in $(Yb,Tm)B_{12}$ solid solutions, using $Tm^{3+}$ ions as sensors [11]. In that case, the effect of Yb ions on the CEF potential felt by Tm impurities turned out to be opposite to that of Ce on $Nd^{3+}$ impurity in the present Ce-based 1-2-10 compounds One possible explanation could be that the contributions from the nearest neighbors, B anions in $YbB_{12}$ and *d*-element cations in 1-2-10, have opposite signs and, therefore, compensation occurs with the hybridization term in the former case, while both contributions add up in the latter. Other possible reasons for this difference could be the "electron-hole symmetry" between Ce and Yb, or the fact that the valence instability involves 2+ ($4f^{14}$) or 4+ ($4f^0$) configurations in Yb and Ce, respectively.

## 2. *Effect of magnetic Nd impurities on the spin gap state*

Experimentally, the introduction of local magnetic moments in a Kondo lattice results in the filling of the gap due to the appearance of quasielastic magnetic scattering associated with Kondo fluctuations, which remains visible down to the lowest temperatures. This is in contrast to pure $CeFe_2Al_{10}$, in which this component appears only when temperature is increased, together with the suppression of the spin-exciton peak. The quasielastic widths, however, are comparable in both cases (Fig. 7). A similar behavior was observed previously in $YbB_{12}$ [11] and can be ascribed to the suppression, as temperature increases, of the singlet ground state specific for Kondo-insulators. In neutron spectra, this results in a transfer of spectral weight from the spin-exciton mode at finite energies to local spin fluctuations, as found in the $CeFe_2Al_{10}$ powder experiments on IN4 (Figs. 4(b) and 7(b), and Ref. [9]). This regime change is also responsible for the maximum observed in the temperature dependence of the static magnetic susceptibility, which denotes the magnetization crossover, near 70 K (temperature of the upturn in the magnetic susceptibility), from Van-Vleck type at low temperature to Curie type at high temperature [33, 17].

In $Ce_{0.85}Nd_{0.15}Fe_2Al_{10}$, the gap suppression due to the appearance of a quasielastic signal was clearly observed. It is worth noting that the high-temperature limit of the quasielastic width does not change considerably with Nd substitution, but the *temperature* at which the variation of the intensity reaches a plateau is reduced. A similar effect (suppression of the spin-gap and strong effect on the spin-exciton) was previously reported for the $YbB_{12}$ with Tm substitution [11].

## 3. *Effect of d-electron doping (Rh substitution)*

Substitution of Rh for Fe in $RE(Fe_{1-x}Rh_x)_2Al_{10}$ (*RE* = La, Ce) results in doping one extra electron on the transition metal sublattice. In the Ce compound, the low-temperature spin gap in the magnetic spectra is suppressed by Rh substitution, starting from the relatively low concentration of $x = 0.05$, as a quasielastic signal gradually develops in the spin-gap region. With increasing Rh concentration, its intensity increases and its width decreases (Fig. 13). Taking the width of the quasielastic line to represent the hybridization strength ($\Gamma_{QE} \sim k_B T_{sf}$), this evolution points to a reduction of hybridization effects originally present in the Kondo-insulator state upon electron doping. Accordingly, a local spin-fluctuation regime is recovered, similar to that found in heavy-fermion compounds [34], with a quasielastic line width at low temperature on the order of 10 K (FWHM).

In a previous work [17], the influence of Rh substitution in CeFe$_2$Al$_{10}$ has been investigated by measuring the magnetic susceptibility, specific heat, electrical resistivity and thermopower. A small concentration of Rh (5% to 20%), results in a magnetic susceptibility enhancement along the orthorhombic *a*-axis, and a Curie-Weiss behavior is observed in a wide temperature range. The low-temperature specific heat is also strongly enhanced by doping, with the linear coefficient at $T = 2$ K shooting up from 0.08 to 0.5 J/mol K$^2$ between 5% and 20% Rh, which provides clear evidence that a heavy-electron metallic state is realized at low temperatures. These results suggest a collapse of the spin and charge gaps due to the weakening of the *c-f* hybridization effect, and are consistent with the neutron spectroscopy results, in which the spin fluctuation energy is seen to reach values typical for heavy-fermion compounds (on the order of 10 K [17]) for Rh concentrations around 20%. It is still an open question to what extent the extra electron provided by Rh, and/or the *d*-band transformation, from 3*d* to 4*d* character, contribute to the reduction of the hybridization strength between Ce *f*-electrons and conduction electrons in this material.

The present situation is at variance with that achieved by isoelectronic substitution of (magnetic) Nd, which suppresses the spin-gap by disrupting the coherence on the RE-sublattice without any effect on the quasielastic linewidth.

## 4. *Spin-exciton mode*

The existence of a spin-exciton mode below the energy of the spin-gap in CeFe$_2$Al$_{10}$ is a unique spectral feature of the Kondo-insulating state. The present polarized-neutron study provides further experimental information by characterizing the mode dispersion along a reciprocal space direction not investigated previously [10]. It confirms that the minimum of the dispersion takes place at the zone boundary point (0,3,0), corresponding to the AF ***Q*** vector describing the AF ordered structure in the compound CeRu$_2$Al$_{10}$ [7]. The energy dispersion is very similar to those measured along the other two directions (Fig. 4) indicating that the strength of the exchange interactions is comparable for all three directions. The intensity falls down rapidly when the reduced wave vector, measured from the AF point, exceeds 0.3 r.l.u., a behavior already noted in CeRu$_2$Al$_{10}$ [7] for the same direction.

The results for CeFe$_2$Al$_{10}$ also reveal that the magnetic scattering intensity of the dispersive mode is dominated by fluctuations of the Ce moment components parallel to the *a* direction (Fig. 2). This likely reflects the strong orthorhombic single-ion anisotropy with an easy *a* axis, as evidenced by static susceptibility results above the temperature of the spin-gap formation. It is interesting to compare this situation with that reported previously for CeRu$_2$Al$_{10}$ [7]. In the latter case, the dispersive spin excitation was primarily observed in the *antiferromagnetic* state, where the Ce moments turned out to be oriented along the (hard) *c* axis. Therefore, the spin excitations in CeRu$_2$Al$_{10}$ were naturally interpreted as *transverse* spin-wave modes. However, we will see below that this may not be the whole story.

It can also be noted that the temperature dependence of the energy dispersion of the spin excitation is different in the two compounds. In CeRu$_2$Al$_{10}$ the energy of the exciton decreases with increasing temperature [7], as expected in the antiferromagnetic magnon model. In the present compound, however, earlier results [10], as well as data from the present work displayed in Fig. 4, indicate a shift to *higher* energies. Simultaneously the mode intensity decreases. Such a behavior was actually predicted in Riseborough's spin-gap excitation model [35, 36] for Kondo insulators. This effect could not be evidenced previously in the "archetype" Kondo insulator YbB$_{12}$ [25] or in SmB$_6$ where the effect of temperature is quite weak [37].

On the other hand, the onset temperature below which the spin excitation develops is clearly higher than $T_N$ in both CeRu$_2$Al$_{10}$ and CeOs$_2$Al$_{10}$, as evidenced from inelastic neutron scattering measurements on the polycrystalline samples and optical conductivity measurements on the single crystalline samples [8]. From the susceptibility data, the onset temperature seems to coincide with the temperature where the susceptibility goes through a maximum [38]. Additional investigations are needed to address this question.

Doping experiments on the Ru-based series Ce(Ru$_{1-x}$Rh$_x$)$_2$Al$_{10}$ ($x$ = 0.05, 0.10) have been reported previously [18]. In that case, substitution on the 4$d$-sublattice adds one extra electron in the $d$-band. The measurements, performed on single-crystals, showed a strong decrease in the intensity of the inelastic mode upon Rh doping, but with only a minor decrease in its excitation energy, and no evidence for the appearance of a sizable quasielastic signal. As a result, the spin-gap remained visible in the solid solutions, at least within the limited concentration range investigated. Most surprisingly, the dramatic suppression of the (supposedly) spin-wave mode was associated with a threefold *increase* in the magnitude of the ordered AF moment (from $\approx$ 0.3 to $\approx$ 1 $\mu_B$) [39, 40]. This observation is difficult to reconcile with the conventional AF magnon picture, as the spin wave intensity varies linearly with moment value (or spin S value), and raises the possibility that the excitation found in CeRu$_2$Al$_{10}$ might *not* arise from the (weak) long-range-ordered magnetic component, but rather from a fluctuating component similar to that occurring in the present Fe-based (paramagnetic) compound.

A salient result of the present study is that, with increasing Rh-concentration in Ce(Fe$_{1-x}$Rh$_x$)$_2$Al$_{10}$ (inset in Fig. 14), the exciton peak shifts to lower energies while transferring spectral weight to the quasielastic component. The quasielastic width gradually decreases to approach values, $\Gamma/2 \sim 1$ meV (Fig. 13), which are typical for heavy-fermion systems as a result of the decrease in the hybridization strength. It is worth noting the correlation, upon Rh doping, between the enhancement of the integrated intensity of the fluctuation spectrum (dominated by the contribution from $M_a$ components) and that of the static magnetic susceptibility, occurring primarily along the $a$ direction [17]. Similar results are known in both CeRu$_2$Al$_{10}$ and CeOs$_2$Al$_{10}$ with respect to bulk susceptibility.

In the IN4 TOF experiments on CeFe$_2$Al$_{10}$ or (Ce,Nd)Fe$_2$Al$_{10}$ powder, it was not possible to study the magnetic exciton in detail because of the localization of its intensity in reciprocal space (resulting in a low spectral weight), of its pronounced energy dispersion, and also because its position is rather close in energy to that of a phonon peak and Nd-ion CEF peak with maximum intensity near 11 meV. The TOF powder measurements were therefore focused on the spin-gap suppression, appearance of a quasielastic signal, and Nd scattering contribution.

On the other hand, the observation of magnetic quasielastic scattering arising from local spin fluctuations is difficult in the TAS measurements with polarization analysis, because the neutron flux is low and the broad and weak magnetic signal is distributed in $Q$ space. In the present study, both types of experiments (IN20 and IN4) thus proved complementary in sorting out the different components of the magnetic spectral intensity, and tracing the spin gap formation at temperatures below 60 K.

**V. CONCLUSION**

We have presented a neutron scattering study of the Kondo-insulator $CeFe_2Al_{10}$ both as a single crystal on IN20 and as polycrystalline powder on IN4. Taken together, the results provide a comprehensive picture of the magnetic excitation spectra in the temperature range from 5 to 80 K and their dependence on several relevant physical parameters. The observed behaviors are ascribed to the different character of the ground states of the Ce ions below and above the crossover temperature ($T^* \sim 70$ K) associated with the spin-gap formation. In particular, the low-energy spectra seem consistent with a Kondo-insulator model taking into account the single-ion anisotropy specific for 1-2-10 systems.

From the foregoing general discussion, it has been argued that hybridization has a strong influence on the CEF potential for intermediate-valence Ce-based systems, resulting in a pronounced decrease of the CEF splitting. In the Kondo-insulator case of $CeFe_2Al_{10}$, the effect is not so strong and it gives rise to some increase in the CEF potential on the $Nd^{3+}$ ion. Such a qualitative difference in behavior for intermetallic compounds all based on $d$-elements ($CeNi_5$ and CeNi vs. $CeFe_2Al_{10}$) is intriguing. The key parameter may be the extent of $f$-electron delocalization in the intermediate valence regime in contrast to hybridization effect in Kondo insulators.

The suppression of the spin gap with Nd substitution in $(Ce,Nd)Fe_2Al_{10}$ is reminiscent of that observed previously on substituting magnetic RE ions in other Kondo-insulator compounds, $(Yb,Tm)B_{12}$ and $(Sm,Gd)B_6$. It may result from the undercompensation induced on the Ce or Yb Kondo lattice by introducing Nd, Gd or Tm trivalent ions with paramagnetic local moments, which disrupt coherence on the periodic Kondo-insulator state. While the Kondo character of the matrix influences the CEF potential sensed by the RE impurity, the impurity paramagnetic moment in turn strongly modifies the ground state of the Kondo insulator.

In $Ce(Fe_{1-x}Rh_x)_2Al_{10}$ the suppression of the spin gap, and appearance of magnetic quasielastic scattering, upon Rh substitution are accompanied by a decrease in the spin-exciton energy. This suggests a continuous transformation taking place as a function of increasing Rh concentration. In this process, $CeFe_2Al_{10}$ system essentially evolves from a Kondo-insulator to a more heavy-fermion like regime, possibly driven by a decrease in the hybridization strength. Our study allows us to trace the evolution of the magnetic excitation spectrum associated with the gradual transformation of a typical Kondo insulator into a dense Kondo system.

As to Rh doping, which was previously reported to enhance the local character of Ce magnetism in $CeRu_2Al_{10}$, it is found here to initiate a transformation of $CeFe_2Al_{10}$ from a mixed-valence Kondo insulator to a heavy-fermion metallic state with a Ce valence state close to integral. This result is in line with the general trend of hybridization change reported for the Ce 1-2-10 series, based on neutron [8], μSR [41] and photo-electron spectroscopy [42] studies, all of which point to a steady decrease in the $c$-$f$ hybridization in the transition metal sequence Fe→Os→Ru.

ACKNOWLEDGEMENTS

The authors wish to acknowledge beam time allocation by the ILL for experiments on IN4 and IN20, DOI: 10.5291/ILL-DATA.4-01-1506, 10.5291/ILL-DATA.4-03-1707, 10.5291/ILL-DATA.4-01-1347, and by the ISIS facility for experiments on MARI and MERLIN, DOI: 10.5286/ISIS.E.RB1610498. We thank J. W. Taylor for his help on MARI, and S. Petit for performing additional spin-wave calculations and for enlightening discussions. We also thank Q. Si, P. Riseborough, D. Yao and S. Acharya for interesting theoretical discussion. This work was partially supported by the Japan Society for the Promotion of Science KAKENHI Grants JP15K05180, JP26400363, JP26800188, JP16H01076, and JP17K05545, JP17K05546. DTA would like to thank JSPS for an Invitation Fellowship.